\documentclass[runningheads]{llncs}
\usepackage{cite}
\usepackage{amsmath,amssymb,amsfonts}
\usepackage{algorithmic}
\usepackage{graphicx}
\usepackage{textcomp}
\usepackage{xcolor}
\begin{document}

\title{Impact  of using a privacy model on smart buildings data for CO2 prediction}

%
%
%

\author{Marlon P. da Silva\inst{1} \and
Henry C. Nunes\inst{1} \and
Charles V. Neu\inst{2} \and
Luana T. Thomas\inst{1} \and
Avelino F. Zorzo\inst{1} \and
Charles Morisset\inst{2}}

\institute{Polytechnic School PUCRS, Porto Alegre, Brazil\\
\and
School of Computing Newcastle University, Newcastle upon Tyne, UK\\
}

\maketitle              



\begin{abstract}
There is a constant trade-off between the utility of the data 
collected and processed by the many systems forming the Internet of Things (IoT) revolution and the privacy concerns of the users living in the spaces hosting these sensors. Privacy models, such as the SITA (Spatial, Identity, Temporal, and Activity) model, can help address this trade-off. In this paper, we focus on the problem of $CO_2$ prediction, which is crucial for health monitoring but can be used to monitor occupancy, which might reveal some private information. 
We apply a number of transformations on a real dataset from a Smart Building to simulate different SITA configurations on the collected data. 
We use the transformed data with multiple Machine Learning (ML) techniques to analyse the performance of the models to predict $CO_{2}$ levels. Our results show that, for different algorithms, different SITA configurations do not make one algorithm perform better or worse than others, compared to the baseline data; also, in our experiments, the temporal dimension was particularly sensitive, with scores decreasing up to $18.9\%$ between the original and the transformed data. The results can be useful to show the effect of different levels of data privacy on the data utility of IoT applications, and  can also help to identify which parameters are more relevant for those systems so that higher privacy settings can be adopted while data utility is still preserved.

\end{abstract}
%




%
\section{Introduction}






The impact of the quality of an indoor environment on the well-being of its occupants is a relatively well-studied problem \cite{arif2016impact}. More than 20 years ago, Redlich \textit{et al.} noted the increase of the Sick-Building Syndrome (SBS), which includes {\em ``upper-respiratory irritative symptoms, headaches, fatigue, and rash"}\cite{redlich1997sick}. Although they deemed $CO_2$ as {\em ``an unlikely cause of SBS"},  a study widely covered in the general press, clearly indicates potential health risks associated with chronic exposure to environmentally relevant elevations in ambient $CO_2$, including {\em ``inflammation, reductions in higher-level cognitive abilities and impact on different body organs"}\cite{jacobson2019direct}. There is, therefore, a clear need for precise and reactive monitoring of indoor $CO_2$, to detect and prevent dangerous situations.

On the one hand, smart buildings deploy IoT architecture usually including $CO_{2}$ and temperature sensors~\cite{8735760}, intended to be used on new services that can be provided, many supported by Machine Learning (ML) techniques. Those services are intended to automatise management and optimise user comfort, security, and safety, quite often with a focus on occupancy measurement ~\cite{labeodan2015occupancy,lam2009occupancy,candanedo2016accurate}. The need for $CO_2$ monitoring is likely to push an increasing deployment of such systems. 

On the other hand, this monitoring faces an increasing privacy concern related to ambient infrastructures. Naieni \textit{et al.} for instance showed that although roughly half of the participants in a survey were comfortable or very comfortable with the collection of presence and temperature data, people nevertheless favour data collection in which they cannot be identified immediately and do not want inferences to be made from otherwise anonymous data~\cite{naeini2017privacy}. 

Lately, new legislation has been introduced to support and regulate personal data  usage and people’s privacy preferences, e.g. GDPR (General Data Protection Regulation)\footnote{https://gdpr.eu/} in the EU/UK and LGPD (Lei Geral de Proteção de Dados)\footnote{https://www.serpro.gov.br/lgpd/menu/a-lgpd/o-que-muda-com-a-lgpd} in Brazil.  A common principle is that of {\em data minimisation}, which specifies that a system should not collect and process more data than needed for its purpose. There is also a clear concern that users must be involved in data collection and processing and their preferences must be considered. As a result, a data protection system must work out a difficult trade-off: minimise data collection to satisfy as much as possible user privacy preference while avoiding a loss to data utility, which might reduce the efficiency of processing, and as a consequence could impact the overall utility of the data to the provided services.


In this paper, we explore this trade-off in the context of a real-world smart building by evaluating how different ML methods perform to predict $CO_{2}$ when different privacy levels are defined. We also evaluate how different levels of privacy impact data utility in comparison to when the whole data is available.

The remainder of this work is organised as follows. Section~\ref{sec:background} presents background on smart buildings, data privacy, and machine learning. Section~\ref{sec:experiment} presents our methodology to configure the dataset and to build our implementation. An experimental evaluation and discussion are presented in Section~\ref{sec:results}. Section~\ref{sec:relatedWork} discusses recent works on privacy in smart spaces and $CO_{2}$ prediction using smart buildings sensor's data. Finally, Section~\ref{sec:conclusion} concludes this work and indicates some future work directions.

\section{Background}
\label{sec:background}
With the revolution of IoT equipment, many smart buildings are emerging, especially in universities and business offices. They are responsible for collecting a huge amount of data from many people every day. In light of this, there is a growing concern about the privacy of these data.

For our research, we collected data from a smart building located at a university in England, applied different settings of the SITA privacy model that will be discussed in the next sections, and used machine learning algorithms to check if there were changes regarding the usability of the data.

Thus, we investigated and collected some information on the topics covered, which are organised as follows in the next sections: Section \ref{sec:smartbuilding} deals with smart buildings, Section \ref{PrivacyModel} presents the SITA  model, and Section \ref{sec:machinelearning} describes some  machine learning algorithms applied in this work.

\subsection{Smart buildings sensors and data}
\label{sec:smartbuilding}

Smart building is a term that has its origins in the larger scenario of building automation, which is the set of practices aimed to improve the control of a building by electronic means. From building automation, emerges the concept of intelligent building, when, in addition to control, we also have historical data enabling us to make predictions\cite{buckman2014smart}.

The drivers for the development of buildings can be said to revolve around adding value to a building\cite{smith2007intelligent}. Reducing energy consumption has now become a driver in its own right, due to increasingly stringent regulations and awareness of climate change. This is recognised in modern buildings as a significant design criterion\cite{ghaffarianhoseini2013sustainable}. In order to achieve these requisites, there are four specific approaches to follow:

\begin{itemize}
    
\item the methods by which building operation information is gathered and
responded to (intelligence);
\item the interaction between the occupants and the building (control);
\item the building's physical form (materials and construction); and
\item the methods by which building use information is collected and used to
improve occupant performance (enterprise).
\end{itemize}

Smart Buildings are Intelligent Buildings, but with additional, integrated aspects of adaptable control, enterprise and materials, and construction. In Smart Buildings, the four methods used to meet the drivers to building progression, mentioned previously, are developed alongside each other, utilising information from one in the operation of another. This is in contrast to Intelligent Buildings, which have largely developed intelligence independently of the other methods.

\subsection{Privacy Model}
\label{PrivacyModel}


In recent years there was an increase in demand for privacy techniques \cite{Seliem:2018} \cite{Sarwar:2021}. This comes in line with an increase in awareness of society to how easily data is collected, distributed, and used in the information age. One consequence is the creation of legislation in different jurisdictions that address this topic, GDPR, LGPD, and PDP\footnote{https://prsindia.org/billtrack/the-personal-data-protection-bill-2019} are a few examples. This legislation tries to organise how data is treated and protected. To tackle the privacy problem the SITA model was developed.

SITA \cite{Schaarup:2013} is a conceptual model that empowers end-users with the ability to control their privacy. It is based on a granular approach to control privacy in applications, and also, uses Maeda's ten laws of simplicity \cite{Maeda:2006} as a design philosophy. As result, an end-user has a simple and intuitive method of controlling how an application can distribute its data. 

The embedded privacy control in most applications works binary, where a user can block the application from sharing all his data or allow it to share all his data in the application. SITA proposes the use of different levels as a way to remediate this. The user can control how much information he is sharing, this, however, comes with a cost. Less information that is shared in the application can degrade an application service because of the lack of precise information. As consequence, a user will need to set the application privacy control in some way so that the application is usable for his needs and also does not share so much information. This trade-off is very common in privacy applications, known as the \textbf{privacy-utility trade-off}. Other frameworks work on a similar premise, allowing more control over the privacy settings \cite{Hess:2002} \cite{Ankur:2009}, however, these models are in general complex, which hinders their widespread usage.

The model is divided into four dimensions: Spatial, referring to the user's location data, such as GPS position, address, and others; Identity, related to the user's personal identification data, such as ID, Name, and Gender; Temporal, date and time information about user activity in the application; Activity, sensitive data about user behaviour, situational data, and preferences. All the data shared with the application developer can be categorised into one of these four groups.

Each dimension can be assigned a level from zero to four. The level represents the amount of privacy for that specific dimension. Where zero represents no access to the data and complete privacy. On the other extreme four represents full access and no privacy protection. The values in-between allow controlling the shared data using aggregation, and obfuscation techniques to granularly control privacy. In that case, the amount of information shared is something between no shared data at all (level zero), and total access to the data (level four). These in-between values are application specific and need to be created by the developer.

An end user can set the level for each dimension, which is called a \textbf{SITA configuration}. The resulting data shared will apply the privacy level for each dimension and share the data with the application. A user can for instance set The SITA level as Spatial two, Identity, three, Temporal zero, and Activity four. The resulting shared data would include Spatial and Identity data modified by some anonymization aggregation, and obfuscation techniques, all the temporal data, and no activity data. In this work to identify different configurations a sequence of four numbers is used. Each number represents the value for one dimension. For example, 4343 means S = 4, I = 3, T = 4, and A = 3.

\subsection{Supervised machine learning algorithms}
\label{sec:machinelearning}

Supervised machine learning algorithms use previously labelled data to train machine learning models. One of its main uses is to allow the inference of further data to a label. There are several algorithms, and variants, that use this method \cite{Jiang:2020}. For this work, we describe the algorithms used in our experiment.

\subsubsection{Linear regression (LR)}

This method is used to predict the value of a variable, named the dependent variable, and based on the value of other variables, named the independent variables. The independent variables are used to compose a linear equation. The dependent variable of a new data entry is predicted using the value of the independent variables as input in the linear equation, resulting in the predicted value.

\subsubsection{Ridge Regression (RR)} 

Multiple regression models, similar to linear regression, create an equation using independent variables to predict a dependent variable. However, unlike linear regression, it creates an equation composed of multiple coefficients. One problem that such an approach suffers is that highly correlated independent variables degrade the model performance. Ridge Regression solves this problem by substituting the least square estimators from multiple-regression models with a ridge regression estimator.

\subsubsection{Random forest (RF)}

Random forest uses multiple decision trees to output a label for a new data entry. The label which is outputted more by the decision trees is the one attributed to the new data entry. Decision trees are multiple decision points in a tree-like format. Each non-leaf node contains a decision that will induce entry to a leaf node. The leaf node contains the label that the data entry will assume. The multiple trees in the random forest are created with some degree of variance. This method reduces the bias and overfitting that using a single decision tree can result in.

\subsubsection{Gradient Boosting Regressor (GBR)}

This model is a variant of the ensemble methods, like RF. In these methods, multiple simple models are combined into a more complex and precise one. For GBR the simpler models usually are decision trees. The result is a more robust model overall. This approach can find any non-linear relationship in the dataset and can treat missing values, and outliers. The main difference between RF and GBR is the process of creating the decision tree and combining the results.

\subsubsection{Decision Tree Regressor (DTR)}

A decision tree is the smaller model that is used in the RF and GBR. It is a classifier in the form of a tree. Each node is a decision and the leaves are the labelled value. A new entry starting at the root is directed at each node to children nodes based on its value. At the leaf node, it receives its label. The tree is built by partitioning the training dataset and building a decision node based on the partitioning.

\section{Experiment}
\label{sec:experiment}

For our experiments, we collected data from a real scenario, the Urban Sciences Building(USB) at Newcastle\footnote{https://api.usb.urbanobservatory.ac.uk}.. In the next sections, we describe the methodology used for collecting and pre-processing the data, applying the privacy model, and the metrics used to measure the performance of the machine learning algorithms.

\label{Experiment}





\subsection{Privacy model for IoT datasets in $CO_{2}$ prediction}


In IoT scenarios, data from sensors can be stored in datasets. A dataset can be useful for history, use in prediction models and pattern detection, forensics, and other cases. However, the dataset can also be used in a malicious way. For example, a malicious data holder can use the data as part of a linkage attack to infringe upon individual privacy. To mitigate this risk one technique is to reduce the precision of data entries in the dataset. This diminishes the data precision, hindering malicious privacy attacks. This approach is used by the SITA model, from an original dataset it generates a private dataset where the original data entries are less precise.

\begin{figure}[!htbp]
    \centering
    \includegraphics[width=0.5\textwidth]{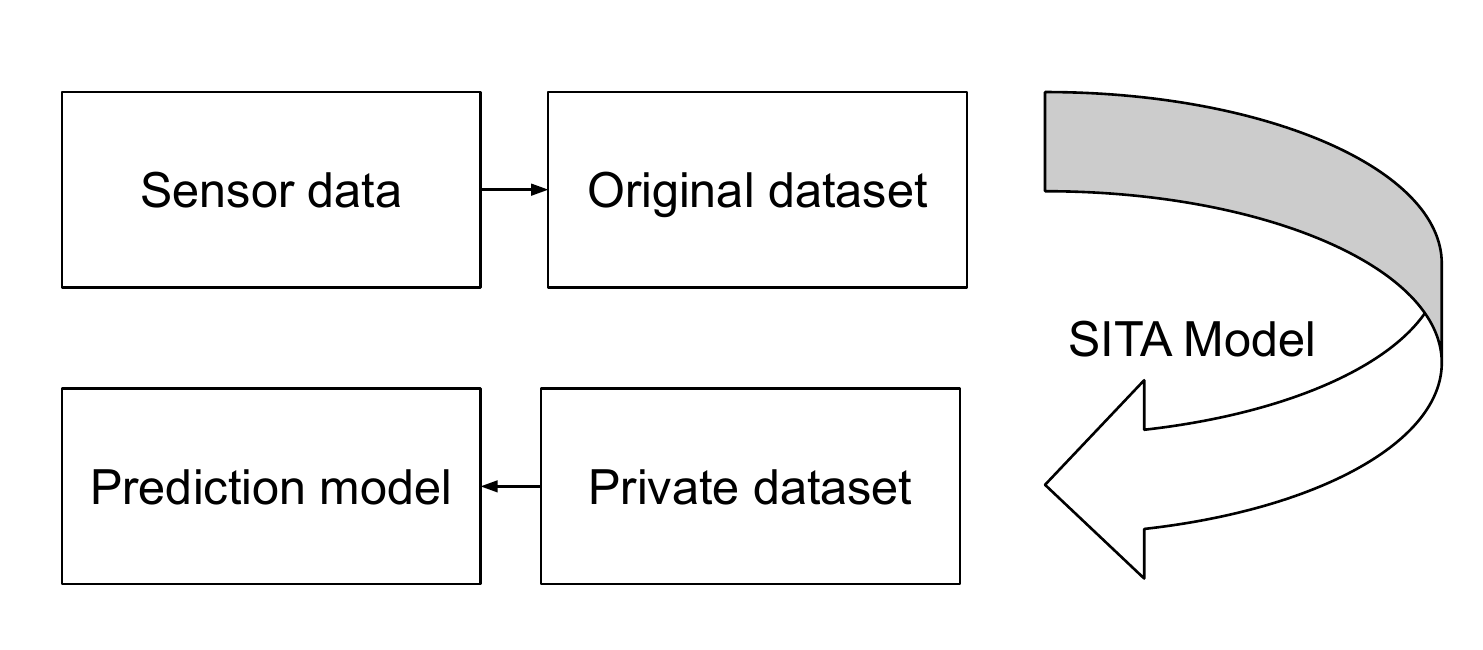}
    \caption{Proposed privacy model used for $CO_{2}$ prediction}
    \label{figura:privacy_model}
\end{figure}

Figure \ref{figura:privacy_model} presents how we apply the SITA model to add different privacy levels to the original data. First, the data is collected from multiple sensors and aggregated in the original dataset. This data will suffer a SITA  transformation based on a SITA Configuration. It is important to note that the same configuration is applied to all the datasets, instead of a configuration for a specific user's data or data entries. The reason for this is that we intend to isolate and analyse the impact of the transformation for each SITA dimension on a machine learning $CO_{2}$ prediction model. The result of the SITA Transformation is a private dataset, a dataset with increased security against privacy attacks. This private dataset is then used to create a prediction model using machine learning techniques.

This approach reflects the possibility that an IoT environment can automatically transform an original dataset, or data as it is added to a dataset using the SITA model. The resulting private dataset can be the only information made available to the data holder. This guarantees an increase in privacy. However, this comes with the cost of data utility, since the modification to the original dataset will decrease its data utility and prejudice the prediction models created from it.

\subsection{Attack model}

Here we present a simple model that can be used to exploit the $CO_2$ readings from a sensor, or its prediction if accurate, to determine which specific person is in a room. There is a number of studies that suggest that people who weigh more produce more $CO_2$ \cite{Magkos:2019}, and there are studies that suggest that males produce more $CO_2$ than females \cite{Yang:2020}. Also, there is the ASHRAE Standard 62.1\cite{ashrae:2007} used to predict $CO_2$ emission inside a building and control air quality, it receives as input the metabolic rate, which is highly based on a person's body composition (fat, muscle, bones, and etc.). In our model, the potential difference of $CO_2$ emission between two individuals will change the $CO_2$ readings of the room, allowing it to be used to identify who is inside a room.

The scenario for this model is a small closed room that is used by just two people. These two people have a significant difference in body composition and are of different sex. We will identify them as Alice (50kg) and Bob (90kg), a third person, Eve, wants to identify who is inside the room without their consent. Eve has some background information, she knows the sex of Alice and Bob and their approximate weight. Eve also has access to the $CO_2$ readings of the room.

\begin{figure}[!htbp]
    \centering
    \includegraphics[width=0.7\textwidth]{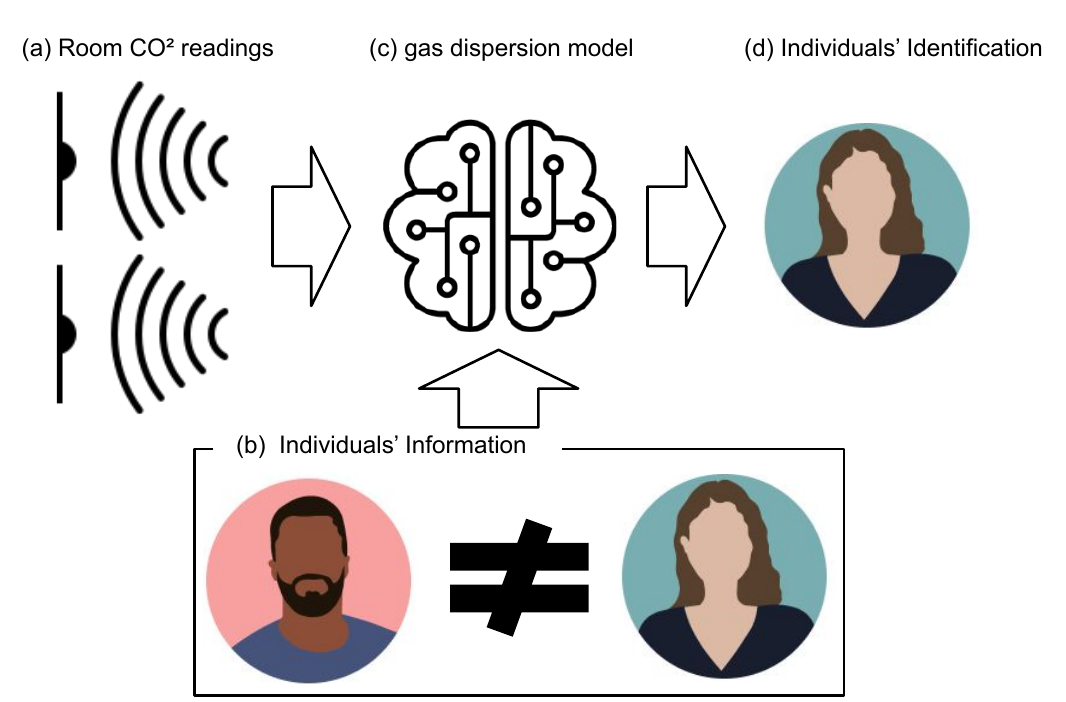}
    \caption{$CO_2$ individuals' privacy attack model}
    \label{figura:attack_model}
\end{figure}

In Figure \ref{figura:attack_model} we summarise our model. The $CO_2$ readings of the room (a), and the background information (b) will be the input of the model of the gas dispersion model (c), that model will as result identify who is inside the room. The gas dispersion model can use the ASHRAE Standard 62.1\cite{ashrae:2007} to predict the $CO_2$ present in the room when it is empty, Alice is in it, Bob is in it, or both. Using this prediction and the actual $CO_2$ readings Eve can then predict who is in the room. The gas dispersion model is not the only option viable for Eve, if she has historic data of the room $CO_2$ she can use an unsupervised machine learning algorithm to cluster the readings in four groups, group 1 when the room is empty, group 2 when Alice is in the room, group 3 when Bob is in the room, and group 4 when both are in the room. Finally, if Eve has a history of $CO_2$ readings and who is present in the room even other supervised machine learning algorithms are viable.

This model is very simplistic and in more complex scenarios can be ineffective. A few possible changes that can impact this model include: Changing the number of people that use the room, this would increase the complexity of isolating the $CO_2$ emission of one person, and more people more complex. People with very similar profiles, same-sex same weight, or very close. A big room would disperse the emitted $CO_2$, and the presence of people inside the room would change very little in the room readings, making it impossible to distinguish who is inside. Ventilation can also disperse the $CO_2$, making it impossible to distinguish who is inside.

\subsection{Description of the Experiment}

The objective of our experiment is to analyse the impact of the SITA privacy model in $CO_{2}$ predictions that use machine learning. More specifically the experiment aims to analyse how changing one dimension of the SITA configuration impacts the performance of the prediction model. Privacy techniques, such as SITA, suffer from the privacy-utility trade-off, increased privacy decreases the data utility which will impact the prediction model. 

We briefly summarise the experiment  conducted in  the following sequence of steps and in Figure \ref{Experiment}. More detail will be provided in the next subsections:

\begin{figure}[!htbp]
    \centering
    \includegraphics[width=0.5\textwidth]{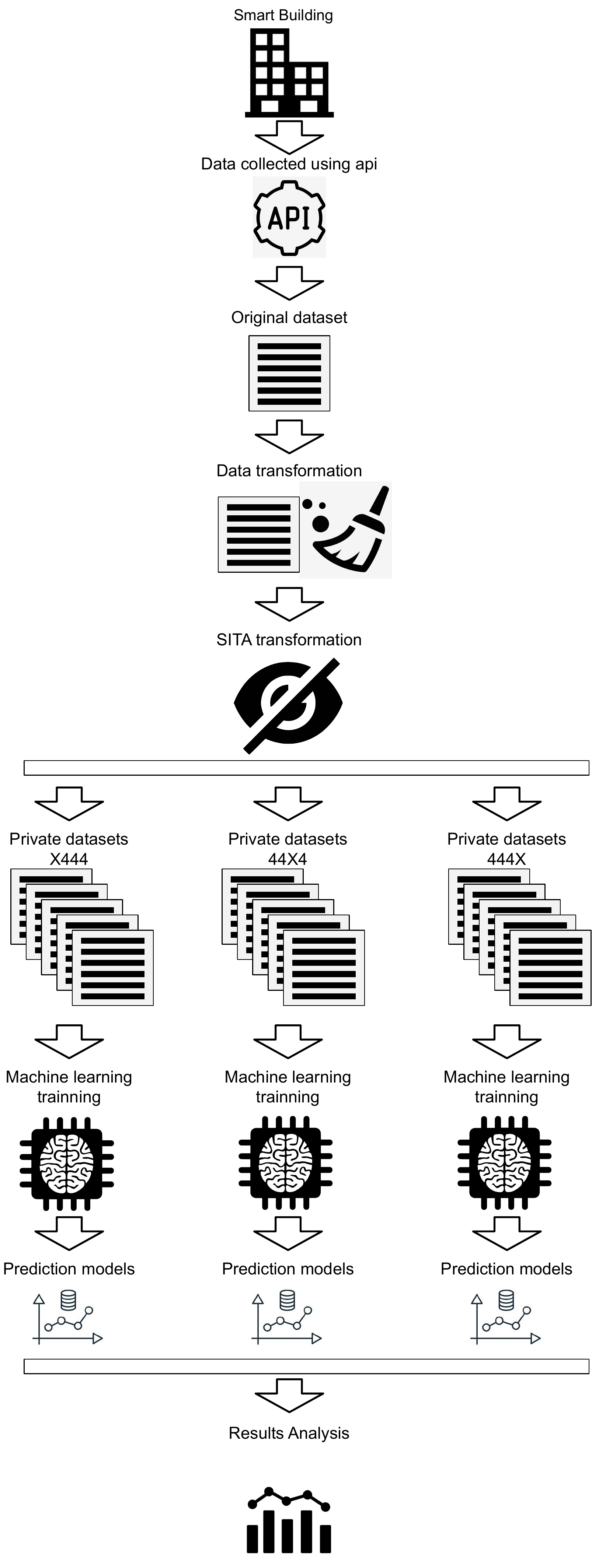}
    \caption{Experiment summary}
    \label{figura:experiment}
\end{figure}

\begin{itemize}
  \item \textbf{Collecting original dataset:} The first step was the collection of the dataset for the prediction model. We used data publicly available from the USB.
  \item \textbf{Data Transformation:} The original dataset was transformed to better suit the SITA transformation.
  \item \textbf{SITA Transformation:} Different datasets were created from different SITA configurations.  We aim to analyze the isolated impact of each SITA dimension. Thus, for each dimension, we changed its level from 0 to 4 while keeping the others in a fixed state. We will analyse the SITA configurations X444, 44X4, and 444X, where X is a number between 0 and 4. This transformation results in five different private datasets for each dimension.
  \item \textbf{ML Training:} Our prediction models are created using LR, RR, RF, GBR, and DTR. The selected techniques are based on the work of Wibisono \textit{et al.} \cite{wibisono2020dataset}. Each technique is used with all datasets generated from the different SITA configurations. 
  \item \textbf{Analysis:} To analyse the impact of a SITA configuration in the prediction model we utilise common ML metrics. The metrics are $R^2$ score, Root Mean Square Error(RMSE), and Mean Absolute Error (MAE). The chosen metrics are also based on the work of Wibisono \textit{et al.} \cite{wibisono2020dataset}. 
\end{itemize}

\subsection{Dataset Scenario and data collection}
\label{Scenario}

Our scenario is composed of the USB.
It is the largest open platform for urban sensing data in Newcastle.
The building includes multiple IoT sensors that can be freely accessed online through their website. An API is also provided, which allows other applications to access the data, in real-time, and also previous data.

The dataset was extracted using a script developed by our team, collecting data from October 2018 to March 2020. We collected data from five sensors: humidity, temperature, occupancy, brightness, and $CO_{2}$. The data is also organised by rooms, with the readings of the sensors in each of these rooms. The rooms differ in size, sensors available, and usage. The historical data is stored in the API by sensors, in different JSON files. Therefore, we had to consolidate all the data into a single file and remove all records that contained at least one missing data.

\subsection{Data Transformation}

To remove all the outliers from our dataset, we set a range of values for each feature. So we have $CO_{2}$ values ranging from 0 ppm to 1,000 ppm (ASHRAE limit for healthy environments\footnote{https://www.ashrae.org/about/position-documents}) and temperature values ranging from 0°C to 50°C. The relative humidity values ranged from 0\% to 100\% and the brightness values ranged from 0 lm to 2000 lm.

After completing the data cleaning as described above, we have a new dataset with about 200,000 records. This dataset is ready to be used in the SITA transformation and in future works.

\subsection{SITA Data Transformation}

Each level of the SITA parameters corresponds to a specific operation on those variables related to that parameters. Below we describe those relationships, and the transformations performed at each level. The Identity dimension is absent in our work, since there is no individual data stored in the datasets, and because of that its respective operations are disabled.

The Spatial dimension is represented by data regarding the room and the zone of each entry. Given a sample input \textit{G.024, 2}, the operations follow:

\begin{itemize}
    \item Level 0: all data deleted. Output: \textit{deleted,deleted}
    \item Level 1: only the general location is given. Output: \textit{building,deleted}
    \item Level 2: only the ground of each room is given. Output: \textit{Ground Floor,deleted}
    \item Level 3: returns full information about the room, omitting the zone data. Output: \textit{G.024, deleted}
    \item Level 4: no transformations are applied. Output: \textit{G.024, 2}
    \end{itemize}

For the Temporal dimension, we took as input the datetime parameter. We present the results given a sample input \textit{20181011141735}.

\begin{itemize}
    \item Level 0: all data deleted. Output: \textit{deleted,deleted}
    \item Level 1: year and month, with day fixed at 01. Output: \textit{20181001,deleted}
    \item Level 2: date. Output: \textit{20181011,deleted}
    \item Level 3: date and hour. Output: \textit{20181011, 140000}
    \item Level 4: no transformations are applied. Output: \textit{20181011, 142735}
    \end{itemize}

For the Activity dimension, we considered the following attributes: CO2, Temperature, Humidity, and Brightness. For the sample input \textit{287.0,27.6,63.8,25.0}, we have the following operations:
\begin{itemize}
    \item Level 0: all data deleted. Output: \textit{deleted,deleted,deleted,deleted}
    \item Level 1: values are rounded up to the two rightmost digits. Output: \textit{300,0,100,0}
    \item Level 2: values are rounded up to the rightmost digit. Output: \textit{290,30,60,30}
    \item Level 3: decimal digits are removed. Output: \textit{287,27,63,25}
    \item Level 4: no transformations are applied. Output: \textit{287.0,27.6,\\63.8,25.0}
\end{itemize}

For the experiment, a SITA configuration is applied to all the entries in the dataset. We applied the following configurations to the original dataset: 4444, 3444, 2444, 1444, 0444, 4434, 4424, 4414, 4404, 4443, 4442, 4441, and 4440. To better organise we will refer to a group of operations that alter the same dimension using X for the dimension it is altering. For example, X444 refers to configurations 4444, 3444, 2444, 1444, and 0444. Note that this transformation from the original dataset resulted in multiple private datasets, one new dataset from each configuration.

\subsection{Machine Learning training}

The ML models were trained using the Kaggle\footnote{https://www.kaggle.com} platform, in a remote computing environment with 4 CPUs and 16 Gigabytes of RAM. The library used for the training was the scikit-learn\footnote{https://scikit-learn.org} version 1.0.2. Before the training, since the algorithms here studied only work with numerical data, we transformed all textual data into numerical over each dataset. After this, the datasets were split into training/testing in a proportion of 80/20 utilising random sampling, with a random state of 10. Over these, we applied the \textit{KFold()} method from scikit-learn, with ten splits and setting the parameter \textit{shuffle} to \textit{true}, to avoid overfitting the model. Each regressor method was then instantiated using their default implementations; the $R^2$, MAE, and RMSE scores were calculated using the function \textit{cross\_val\_score()}, indicating in the \textit{score} parameter the respective metric.

\section{Results and Discussion}
\label{sec:results}
This section summarises the results of our experiments. We have three sets of SITA configurations, each one varying by one dimension and keeping the other three unmodified (i.e. level 4 of the model). We did this to better comprehend the impact of applying different SITA options over the dataset.

We measured the results accordingly to three metrics: $R^2$ score, Mean Absolute Error (MAE), and Root Mean Squared Error (RMSE). $R^2$ score, also known as the Coefficient of Determination, measures the proportion of the variance in the dependent variable that is predictable from the independent variable(s). In other words, it represents the correlation between the predicted outcomes of a model and their real values\cite{draper1998applied}. Mean Absolute Error is the average difference between predicted and real values, representing how much the model misses the expected value. The Root-mean-square error, like MAE, is also an error metric, but here the measurement is calculated by the square root of the average of squared differences between prediction and actual observation. Although similar (with MAE being usually recommended, due to its easier interpretation\cite{willmott2005advantages}), MAE and RMSE exhibit distinct behaviours in specific circumstances, such as with large test sizes; thus, a combination of both metrics are often required to assess model performance\cite{chai2014root}.

For each SITA configuration analysed, we applied ten-fold cross-validation over the trained models, and collected their $R^2$, MAE, and RMSE values at the end of each execution; after all executions were completed, we calculated the average scores of each model/\/configuration pair.
\begin{figure}[!htbp]
    \centering
    \includegraphics[width=0.5\textwidth]{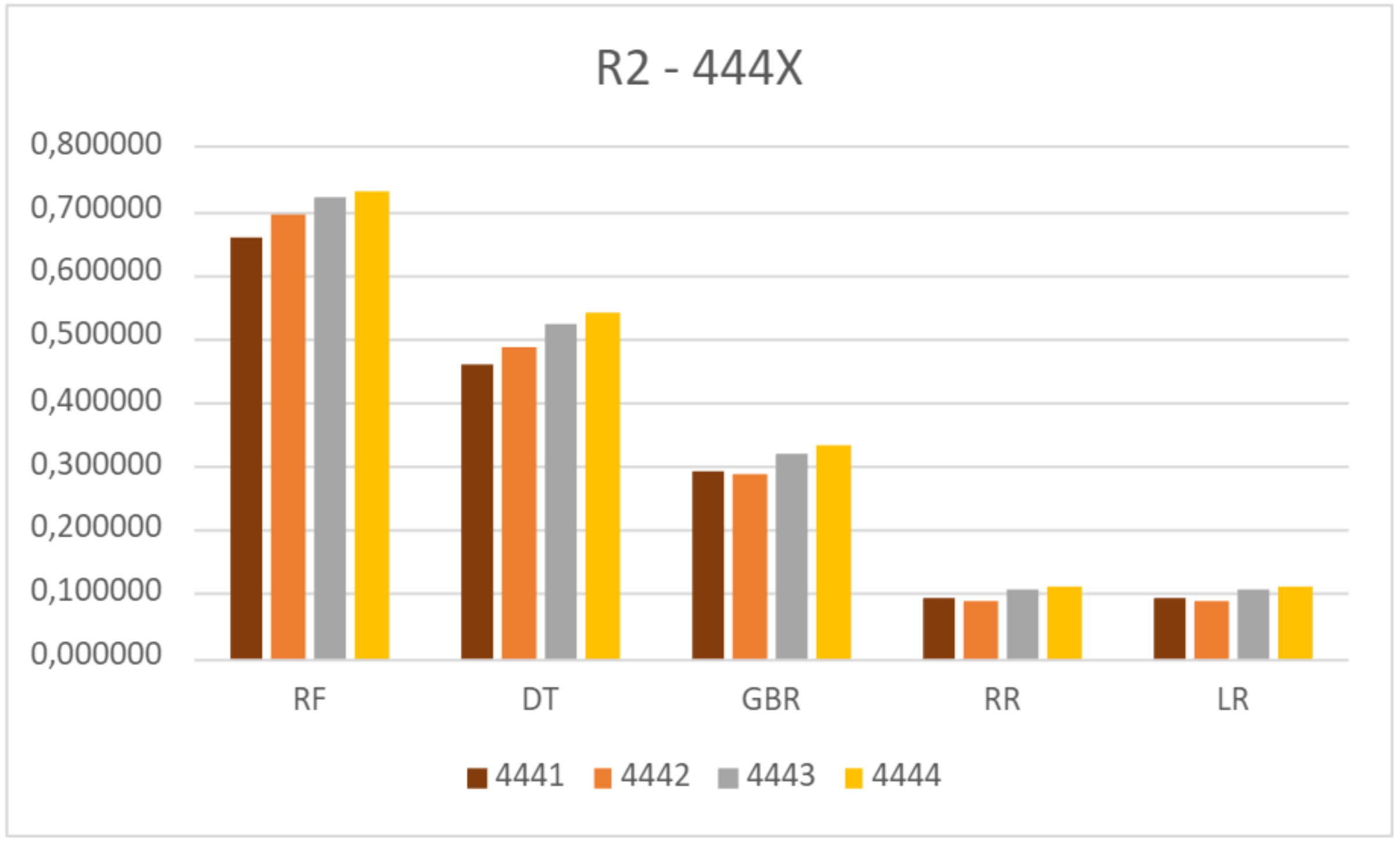}
    \caption{$R^2$ score for machine learning algorithms in 444X SITA dimension.}
    \label{figura:Arq_original}
\end{figure}

\begin{figure}[!htbp]
    \centering
    \includegraphics[width=0.5\textwidth]{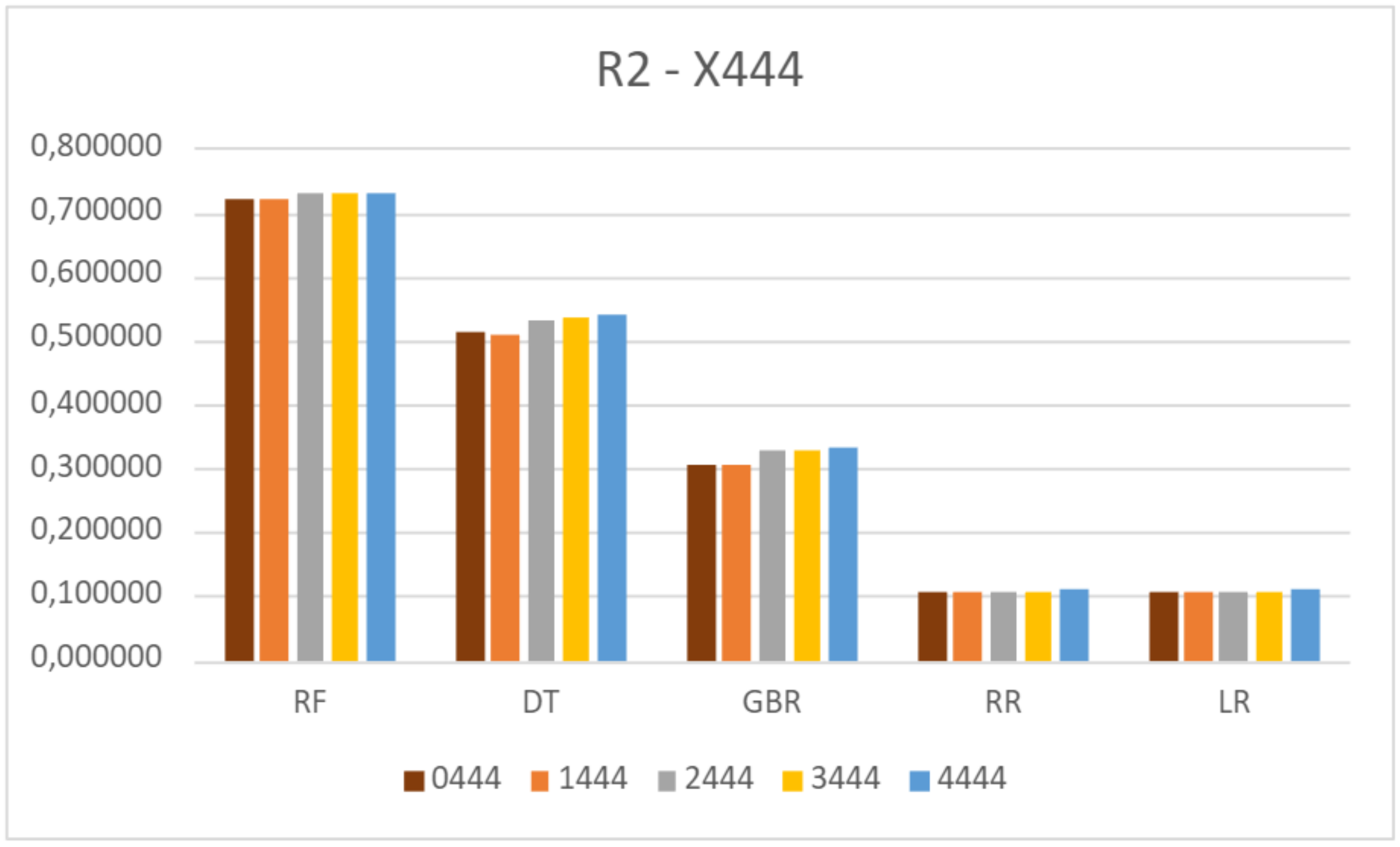}
    \caption{$R^2$ score for machine learning algorithms in X444 SITA dimension.}
    \label{figura:Arq_original}
\end{figure}

\begin{figure}[!htbp]
    \centering
    \includegraphics[width=0.5\textwidth]{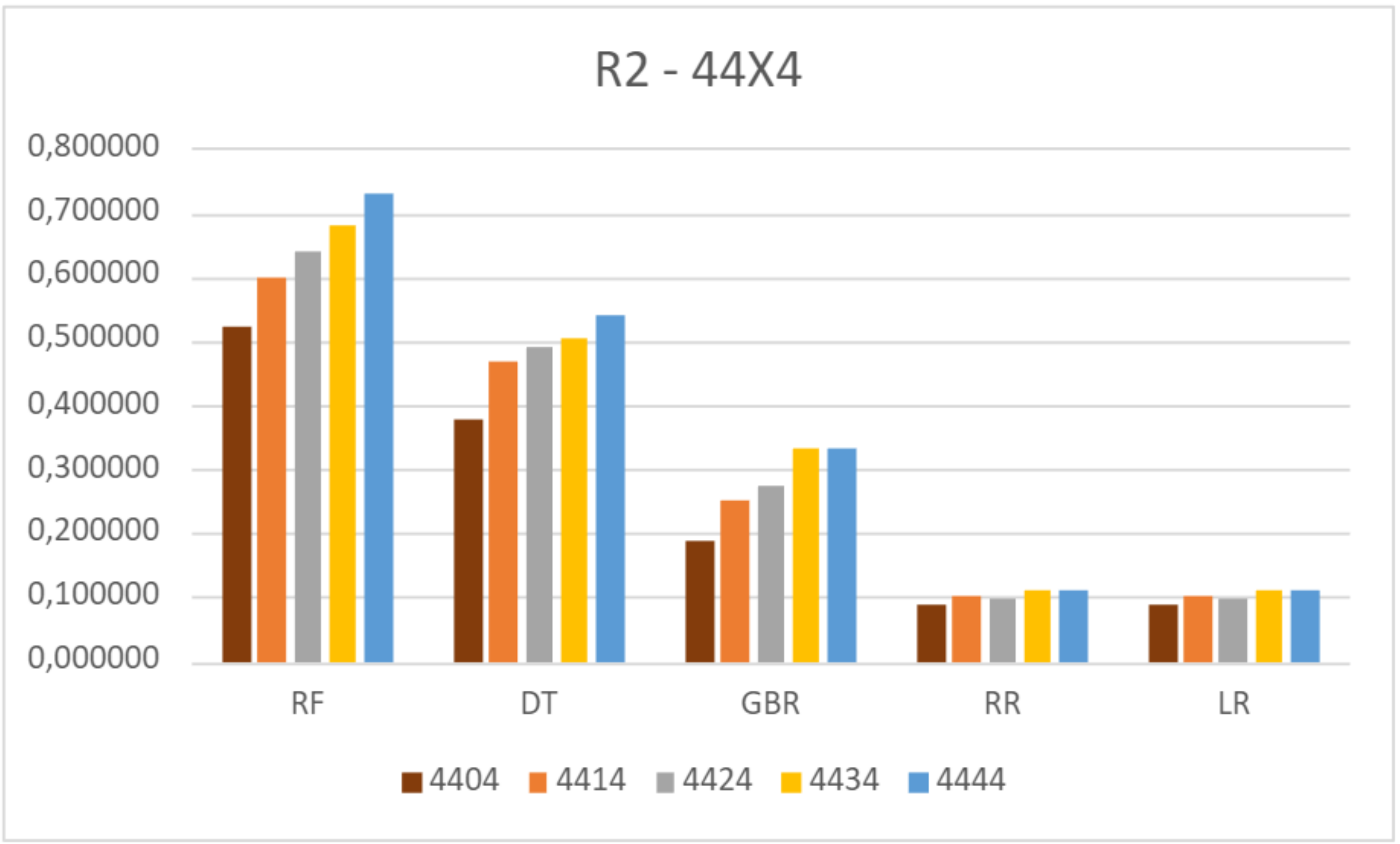}
    \caption{$R^2$ score for machine learning algorithms in 44X4 SITA dimension.}
    \label{figura:Arq_original}
\end{figure}

\begin{figure}[!htbp]
    \centering
    \includegraphics[width=0.5\textwidth]{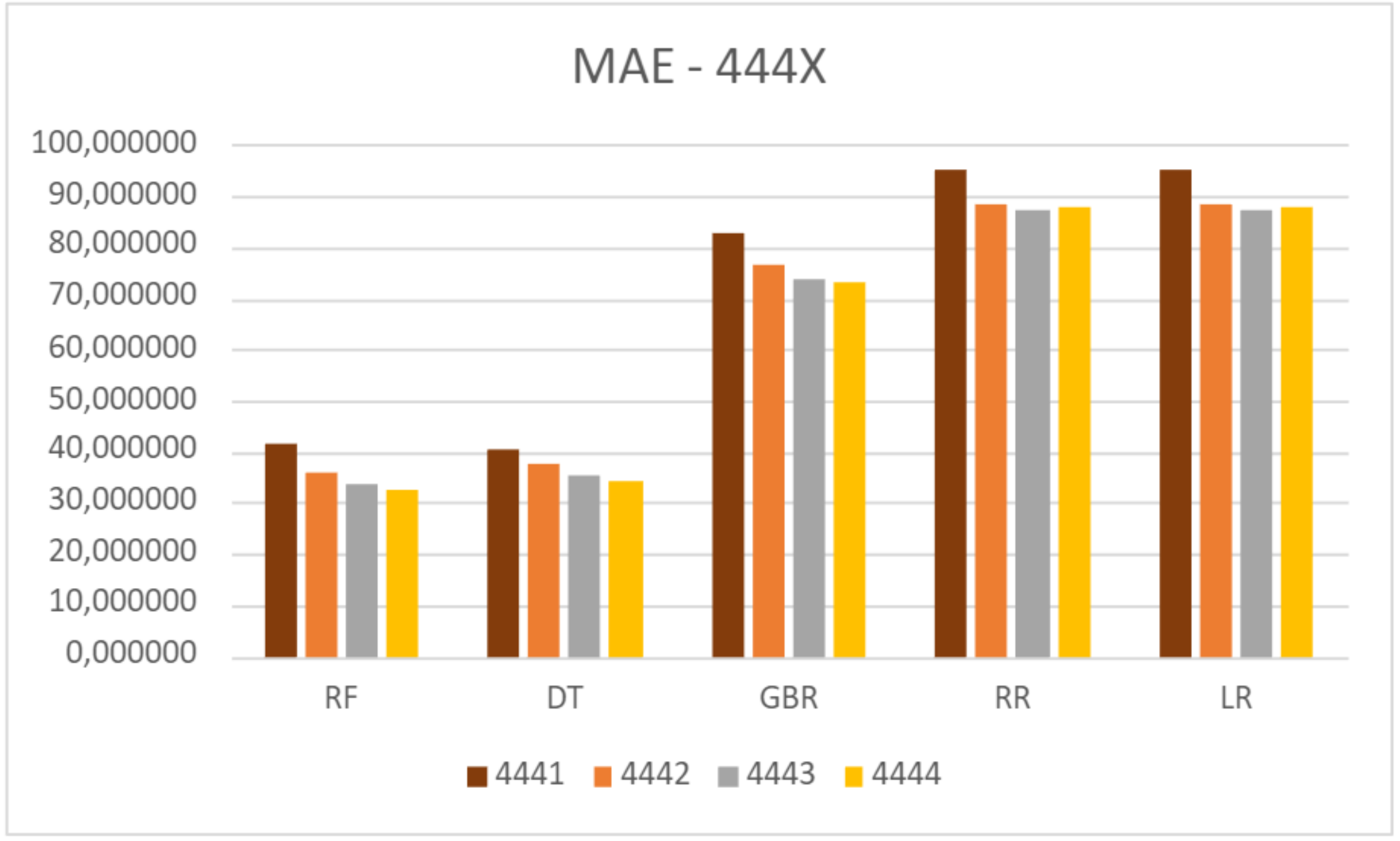}
    \caption{MAE score for machine learning algorithms in 444X SITA dimension.}
    \label{figura:Arq_original}
\end{figure}

\begin{figure}[!htbp]
    \centering
    \includegraphics[width=0.5\textwidth]{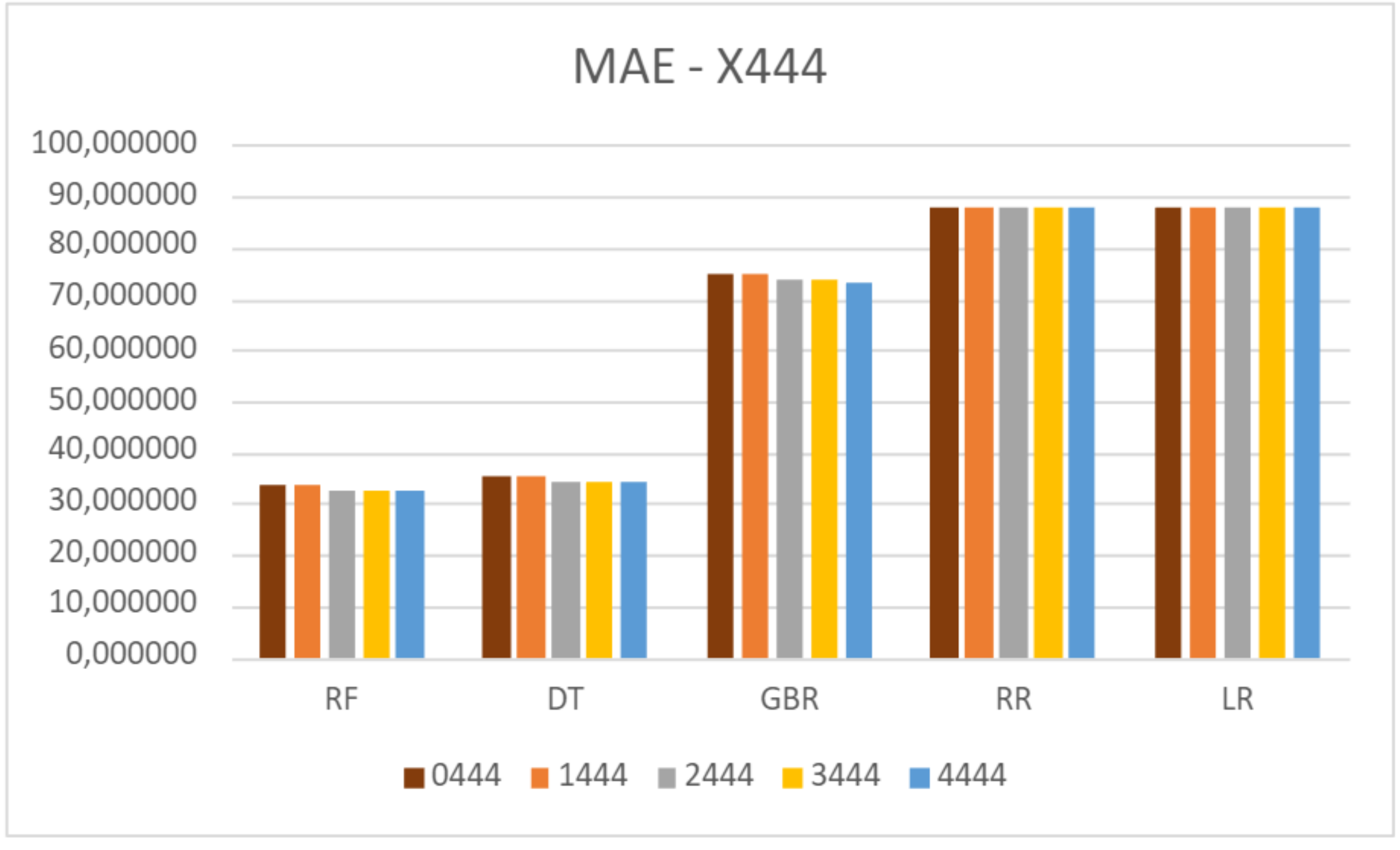}
    \caption{MAE score for machine learning algorithms in X444 SITA dimension.}
    \label{figura:Arq_original}
\end{figure}

\begin{figure}[!htbp]
    \centering
    \includegraphics[width=0.5\textwidth]{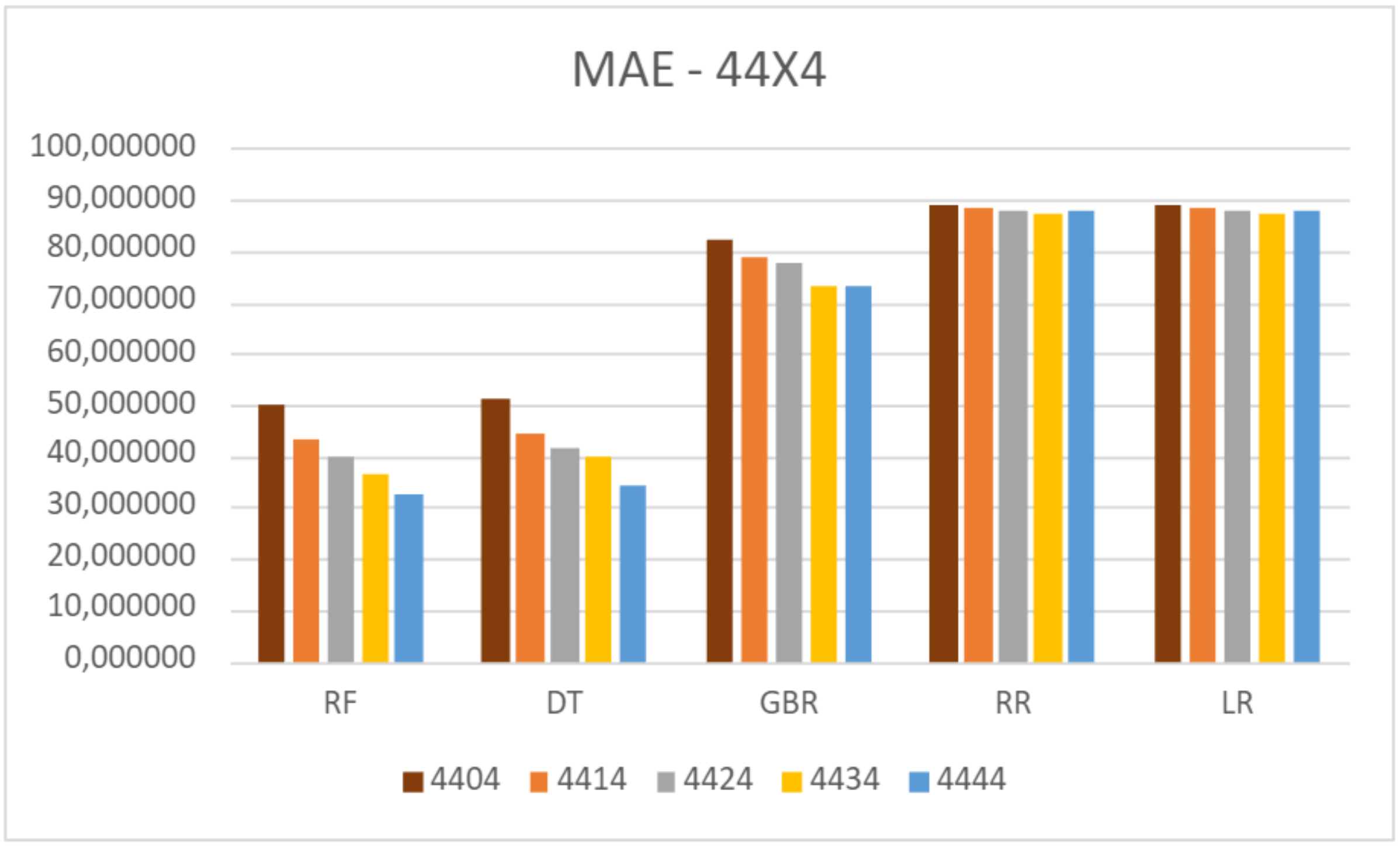}
    \caption{MAE score for machine learning algorithms in 44X4 SITA dimension.}
    \label{figura:Arq_original}
\end{figure}

\begin{figure}[!htbp]
    \centering
    \includegraphics[width=0.5\textwidth]{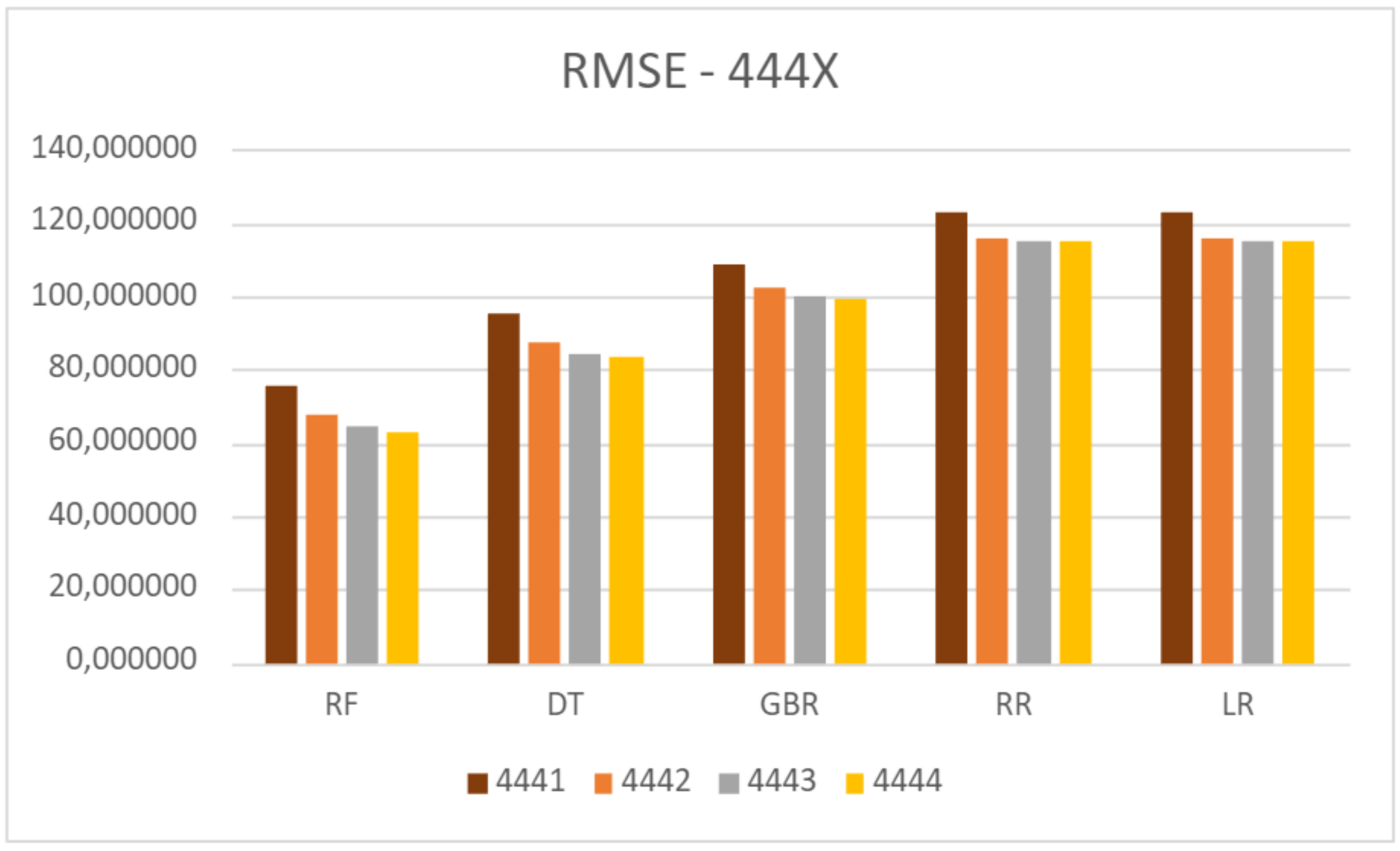}
    \caption{RMSE score for machine learning algorithms in 444X SITA dimension.}
    \label{figura:Arq_original}
\end{figure}

\begin{figure}[!htbp]
    \centering
    \includegraphics[width=0.5\textwidth]{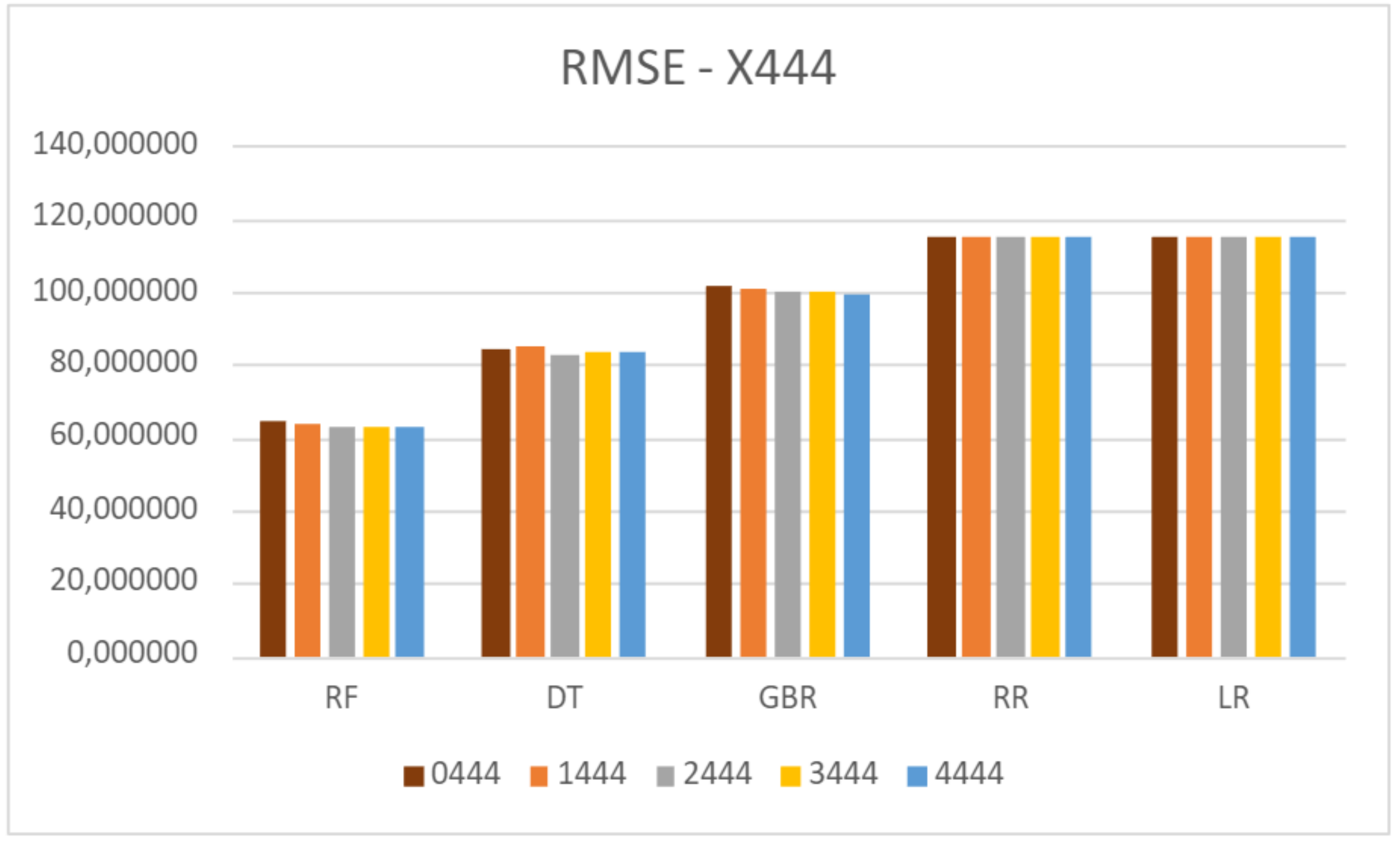}
    \caption{RMSE score for machine learning algorithms in X444 SITA dimension.}
    \label{figura:Arq_original}
\end{figure}

\begin{figure}[!htbp]
    \centering
    \includegraphics[width=0.5\textwidth]{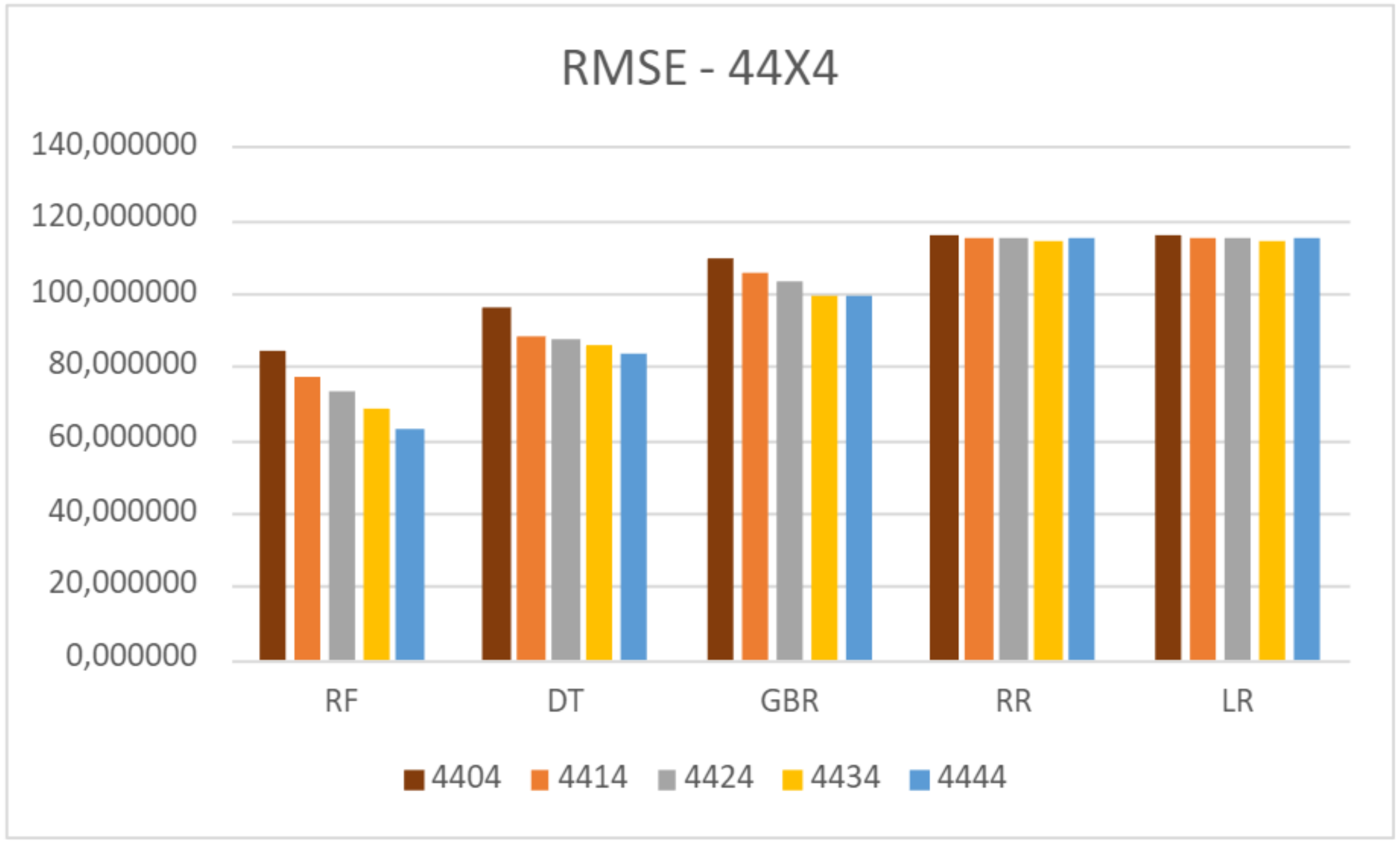}
    \caption{RMSE score for machine learning algorithms in 44X4 SITA dimension.}
    \label{figura:Arq_original}
\end{figure}




For every specific setting, we ran all five algorithms, and measured the results according to three parameters: the coefficient of determination ($R^2$ score - figures 2 to 4), the Mean Absolute Error (MAE - figures 5 to 7), and the Root Mean Squared Error (RMSE - figures 8 to 10). Since there are significant performance differences between Random Forest and Decision Tree models compared to the other algorithms evaluated, we will focus our analysis on these first two.

Regarding the $R^2$ score, the Random Forest algorithm outperforms all other approaches analysed, with an average value of 74.29\% for the baseline data. When taking the Activity dimension as variable, the minimum average score of this model is 66.05\%. This represents a performance decrease of 9.44\% regarding to the baseline. Even in this worst-case scenario, the Random Forest model still outperforms the second-best algorithm (Decision Tree) by 23.62\%, even when considering the baseline for the latter. It is important to observe that we did not apply level 0 of SITA transformations in that dimension since this operation would erase all CO2 data, thus making it impossible to predict its levels.

For the Temporal dimension, the minimum average score of Random Forest is 51.80\% for the Temporal privacy setting of 0 (i.e. all data deleted). This represents a performance decrease of 28.98\% regarding the baseline. In this worst-case scenario for the Random Forest model, the Decision Tree baseline performance is 3.05\% superior, with all other temporal settings of the Random Forest performing better than the Decision Tree baseline, and all other temporal settings of the Decision Tree underperforming the Random Forest worst-case.

When considering the Spatial dimension, the Random forest model produces an average $R^2$ score of 71.86\% when the privacy setting is at 0, a decrease of 1.73\%. This value is 33.00\% higher than the Decision Tree score for the baseline case, having this model in the same privacy level a score of 51.16\%, being 28.81\% lower than RF.  

Analysing the Mean Absolute Error, the lowest score for the baseline was obtained running the Random Forest model, with a result of 32,40\%. With the Activity dimension as variable, the maximum average score of such algorithm equals to 41,28\% when privacy level = 1 for Activity. This represents an error increase of 27,38\% regarding to the baseline. The Decision Tree algorithm produces an MAE value of 40,60\%, which represents a decrease of 1,64\% in relation to the Random Forest.

For the Temporal dimension, we have a maximum average score of 49,85\% (privacy level = 0), 53,50\% higher than the baseline. The same configuration when applied to Decision Tree produces an MAE score of 50,93\%, 2,17\% higher than the Random Forest model.

The Spatial dimension has an MAE score of 33,63\% on the strictest privacy level for the RF model. This represents an increase of 3,56\%, considering the baseline. The Decision Tree model, conversely, has an MAE score of 35,49\%, an increase of 5,53\% from the RF.

Finally, we look at the RMSE metric. The lowest value was obtained with the Random Forest model, with a score of 62.99\%. For the Activity dimension, the highest value was 75,47\%, 13.83\% higher than the baseline. When we set Temporal and Spatial domains as variables, their respective scores were 84.20\% (an increase of 33.67\%) and 64.31\% (2.09\% higher than the baseline). 

\subsection{Discussion}

Firstly, our results confirm the experimental data presented in \cite{wibisono2020dataset}, regarding the performance of the ML algorithms then analysed. Linear Regression and Ridge Regression, being fairly simple algorithms 
, are expected to perform poorly than more sophisticated methods, especially on large datasets. The other algorithms used in our experiments can be seen as belonging to the same family, with Decision Tree being the basis for both Random Forest and Gradient Boosting. However, some careful tuning is necessary for the latter to achieve good results,  which makes it harder to apply the method over different domains. 
Regarding Decision Tree and Random Forest methods, since the latter is an averaging of multiple instances of the first (thus mitigating possible errors due to overfitting), the results of our experiments confirm the expected performances. 
We also show that RF is the only algorithm between those analysed that produces $R^2$ scores over 70\%. 
When analyzing the Mean Absolute Error, our results show that the performances of RF and DT algorithms are very close, but the RMSE values present a more significant difference between these two methods. This can be explained by the fact that RMSE has a tendency to be increasingly larger than MAE as the test sample size increases, thus exacerbating small differences in MAE values between the two approaches.

Another discussion can be made about the impact of applying our SITA implementation over the chosen dataset. Our results show that the dimensions present different sensibilities to more restrictive privacy settings. Taking the $R^2$ score, we show that the Spatial dimension is the least affected, and the Temporal dimension the most affected, with Activity being in an intermediate place. This can be used to better understand the importance of different variables in applying ML techniques. Also, by analysing the scores of each privacy setting, we observe that the Activity dimension has a score below 70\% when the privacy setting is lower than 3; the same occurs for the Temporal dimension with privacy setting lower than 4 (reflecting the higher sensitivity of this dimension), and it is not observed in the Spatial dimension in any configuration. with this we demonstrate that it is possible, through different SITA settings, to improve the users' privacy and keep ML services functional.


An interesting approach for further research on this topic is the use of other machine learning algorithms, including more powerful techniques such as deep learning. Exploring other domains such as healthcare, social media, and other IoT scenarios for example are also interesting further directions.

\section{Related Work}
\label{sec:relatedWork}
There are numerous works related to the prediction of $CO_{2}$ in IoT environments using machine learning algorithms. The $CO_{2}$ monitoring is an important component of controlling the air quality of a room, which when correctly managed provides well-being, controls general air pollution, and detects potential harms, such as fire. Creating a prediction model can be positive in cases presented by Kapoor \textit{et al.} \cite{Kapoor:2022} where smart sensors are not available, also a model can be used to help in the building design. In his work, they present a model working with multiple machine learning algorithms and achieve a precise model.

Other works are developed in a similar fashion using machine learning algorithms in an IoT scenario to create a prediction model for $CO_{2}$. Vanus \textit{et al.} \cite{Vanus:2016} use the value of other sensors like temperature and humidity to predict the $CO_{2}$ value in a Smart Home scenario. In another study, Sharma \textit{et al.} \cite{Sharma:2018} describe the building of a sensor network to detect different pollutant gases beyond $CO_{2}$, although still in development a model is described. An artificial neural network is used to predict air quality and to fully control an IoT network, including air-conditioning, and ventilation based on the work of Tagliabue \textit{et al.} \cite{Tagliabue:2021}. 

There are other works in the prediction of air quality and $CO_{2}$ \cite{Vanus:2019} \cite{Kadam:2018} \cite{Khorram:2019}. However, the use of such prediction models and privacy models as the one presented in Section \ref{PrivacyModel} are not common. The readings of a $CO_{2}$ with other background information can be used in a linkage attack \cite{Merener:2012} in the original dataset, or real-time data, to discover information about individuals, for example, who was present in a room, patterns of movements in a building, among others.

\section{Final consideration and future work}
\label{sec:conclusion}


In this work, we analysed the trade-off between privacy and utility for $CO_{2}$ prediction on a real dataset in the context of smart buildings. Therefore, several transformations were implemented on the original data to simulate different privacy levels and generate new transformed datasets that were used as input to train five distinct machine learning models for $CO_{2}$ prediction. 

The results show that the performance of Regression based machine learning techniques is lower than decision Tree-based techniques. The use of the privacy model, as expected, deteriorated the performance of all algorithms. More aggressive SITA configurations resulted in worse  performance and each dimension has a different impact on the prediction models. The highest impact was observed when higher privacy levels were simulated on the Temporal dimension.

As future research directions, our model could be improved by using  \textbf{Syntactic Anonymity} \cite{sweeney:2002b} with SITA to increase even more the data privacy. To the best of our knowledge, there is no work of this kind yet. Also, the inclusion of \textbf{Differential Privacy} \cite{dwork:2008} is another possibility that could improve even more the privacy model, since it is a more powerful privacy definition than syntactic anonymity.

\bibliographystyle{splncs04}
\bibliography{ref}

\end{document}